\pdfoutput=1

\documentclass[aps,showpacs,nofootinbib,floatfix]{revtex4}

\usepackage{graphicx}

\newcommand{\be}{\begin{equation}}
\newcommand{\ee}{\end{equation}}
\newcommand{\ben}{\begin{eqnarray}}
\newcommand{\een}{\end{eqnarray}}

\newcommand{\la}{{\lambda}}

\newcommand{\cL}{{\cal L}}

\newcommand{\na}{\nabla}

\newcommand{\tpe}{{\tilde p}}

\newcommand{\hI}{\hat I}
\newcommand{\hg}{\hat g}
\newcommand{\hR}{\hat R}
\newcommand{\hna}{\hat \nabla}

\newcommand{\tT}{\tilde T}

\newcommand{\zpsi}{\psi^{\ast}}

\begin{document}

\title{Collapse of Charged Scalar Field in Dilaton Gravity }


\author{Anna Borkowska and Marek Rogatko}
\affiliation{Institute of Physics \protect \\
Maria Curie-Sklodowska University \protect \\
20-031 Lublin, pl.~Marii Curie-Sklodowskiej 1, Poland \protect \\
aborkow@kft.umcs.lublin.pl \protect \\
rogat@kft.umcs.lublin.pl \protect \\
marek.rogatko@poczta.umcs.lublin.pl}

\author{ Rafa{\l} Moderski}
\affiliation{Nicolaus Copernicus Astronomical Center \protect \\
Polish Academy of Sciences \protect \\
00-716 Warsaw, Bartycka 18, Poland \protect \\
moderski@camk.edu.pl }

\date{\today}

\begin{abstract}
We elaborated the gravitational collapse of a self-gravitating complex charged scalar field
in the context of the low-energy limit of the string theory, the so-called dilaton gravity.
We begin with the regular spacetime and follow the evolution through the formation of
an apparent horizon and the final central singularity.
\end{abstract}

\maketitle

\section{Introduction}
The long-standing prediction of general relativity is the occurrence 
of spacetime singularity inside black holes. The singularity theorems
of Penrose and Hawking \cite{ph}
predict the occurrence of spacetime singularities 
inside black holes under very plausible assumptions, but tell nothing  about
the geometrical and physical nature and properties of the emerging singularities.
\par
Until recently the only known generic singularity was the 
Belinsky-Khalatnikov-Lifshitz (BKL) one \cite{bkl}. 
According to this picture spacetime 
develops a succession of Kasner epochs in which the axes  of contraction
and expansion change chaotically. This singularity has a strong oscillatory 
character, which is highly destructive for any physical object. Soon after
Belinsky {\it et al.} \cite{b}
used a scalar field to have an insight into a cosmological 
singularity problem of BKL. They found that scalar fields destroyed
BKL  oscillations and singularity became monotonic.
Recently, it was shown \cite{dam} that the general solutions near
spacelike singularity in superstring theories and in M-theory 
(the Einstein-dilaton-p-form field) exhibit
an oscillatory character of BKL type.
\par
For a Schwarzschild black
hole it has been shown that the asymptotic portion of spacetime near
singularity is free of aspherical perturbations propagated from the
star's surface since the gravitational radiation is infinitely diluted
as it reaches the singularity. On the other hand, the internal structure
of Reissner-Nordstr\"om (RN) or Kerr black hole differs significantly from
the above picture. The singularity becomes timelike and both of
these spacetimes possess Cauchy horizons (null hypersurfaces beyond
which predictability breaks down). In the last few years a new picture
of the inner structure of black holes was achieved, according to which
the Cauchy horizon inside RN or Kerr black hole transforms into a null,
weak singularity, i.e., an infalling observer hitting this null
singularity experiences only a finite tidal deformation \cite{ori},
\cite{br}.
The curvature scalars and mass parameter diverge along this Cauchy singularity and this
phenomenon is known as mass-inflation.
The physical mechanism on which the Cauchy horizon singularity
is based
is strictly
connected with small perturbations (remnants of gravitational collapse)
which are gravitationally blueshifted as they propagate in the black hole
interior parallel to the Cauchy horizon. For a toy model of spherically
symmetric charged black hole the main features of singularity at
the inner horizon were first deduced analytically from simplified examples
based on null fluids \cite{hi}-\cite{is}.
\par
Most of the conclusions 
supporting the existence of a null weak Cauchy horizon singularity
were obtained by means of a perturbative
analysis. The full non-linear investigations of the inner structure of
black holes were given in Ref.\cite{gg} where the authors
revealed that a central spacelike singularity is located deep inside a
charged black hole coupled to a neutral scalar field. The existence of a
null mass-inflation singularity was established by Brady and Smith 
\cite{br},
in their studies of non-linear evolution of 
a neutral scalar field on a
spherical charged black hole.
Burko \cite{bur} also studied the same problem and found a very 
good agreement
between the non-linear numerical analysis and the predictions of the
perturbative analysis.
Expressions for the divergence rate of the blue-shifted
factors for that model valid everywhere along
Cauchy horizon, were given analytically in Ref.\cite{bo}.
One should have in mind that all these  numerical works were beginning on RN
spacetime and a black hole formation was not calculated.
Numerical studies of the spherically symmetric collapse of a massless scalar field in the
semiclassical approximation were conducted in Ref.\cite{aya97}.
Piran {\it et al.} \cite{hod98} studied the inner structure of a charged 
black hole formed during the gravitational collapse of a self-gravitating charged 
scalar field.
They started with a regular spacetime and conducted the evolution
through the formation of an apparent horizon, Cauchy horizon and a final
singularity. The results obtained in \cite{hod98} were confirmed and refined in Refs.
\cite{ore03,han05}.
\par
The effect of pair creation in the strong electric fields in a dynamical model of a collapse
of the self-gravitating electrically charged massless scalar field was elaborated in \cite{sor01}.
The authors studied the discharge below the event horizon and its influence on the dynamical
formation of the Cauchy horizon. On the other hand, the dynamical formation and evaporation
of a spherically charged black hole supposed initially to be non-extremal but tending towards
the extremal black hole and moreover emitting Hawking radiation was investigated in Ref.\cite{sor01b}.
Recently, a spherically symmetric charged black hole with a complex scalar field, gauge field and 
renormalized energy momentum tensor (in order to take into account the Hawking radiation)
was considered in \cite{hon10}.
When the Hawking radiation was included it turned out that the inner horizon was separated from
the Cauchy one. Studying the neutralization of the charged black hole in question
it was found that the inner horizon evolved into a spacelike singularity. Using the exponentially
large number of scalar particles it happened that one could extend investigations inside the inner horizon.
Recently, the response of the Brans-Dicke field during gravitational collapse of matter was analyzed \cite{hwa10}
while the internal structure of charged black hole including Hawking radiation and discharge was elaborated
in \cite{hwa10a}.
In Ref.\cite{sor10} the first axisymmetric numerical code testing the gravitational collapse of 
a complex scalar field was presented.
Also the non-linear processes with the participation of an exotic scalar field modeled as a free scalar field
with an opposite sign in the energy-momentum tensor, were considered \cite{dor10} due to the case when
RN black hole was irradiated by this kind of matter.
\par
In this paper we shall consider the implication of superstring
gravity for the dynamical collapse of charged complex scalar field. 
The famous Wheeler's dictum that {\it black holes have no hair} predicts
that the inner structure of the black hole will not depend on the collapsing
fields. In mathematical formulation this conjecture corresponds to the so-called
black hole uniqueness theorem \cite{uniq} (classification of the domains of outer communication of regular
black hole spacetimes). On the other hand, various aspects of uniqueness theorem for four-dimensional
black holes in the low-energy string theory was widely treated in Refs.\cite{len}.
\par
In previous papers
\cite{fiel} we have examined the intermediate and late-time behaviour of matter fields (scalar and fermions)
in the background of dilaton black holes. In some way the present studies will generalize the former ones
to the more realistic toy model of the dynamical collapse. In what follows
we assume
that the considered Lagrangian for the charged complex scalar field will be
coupled to the dilaton via an arbitrary coupling, i.e.,
$e^{2 \alpha \phi} \cL (\psi, \zpsi, A)$, in the {\it string frame}.\\ 
The outline of the remainder of the paper is as follows.
In Sec.II we derive the equations describing the collapse of the charged scalar field in the 
presence of  non-trivial coupling to the dilaton field.
Sec.III was devoted to the numerical scheme
applied in our investigations. We discussed the numerical algorithm, an adaptive grid used in computations and
the boundary and initial conditions for the equations of motion for the considered problem. We also paid attention
to the accuracy of our numerical code. Sec.IV is assigned to the discussion of the obtained results.
In Sec.V we concluded our researches.

\section{dilaton black hole}
In this section our main interests will concentrate on the behaviour  of
the collapsing complex charged scalar field when gravitational interactions
take a form typical for the low-energy string theory, the so-called dilaton gravity . To take into 
account the unknown coupling of the dilaton field to the considered
charged complex scalar field, we choose the action in the form 
\be
\hI = \int d^{4} x \sqrt{- \hg} \left [ e^{- 2 \phi}
\left (
\hR - 2 \left ( \hna \phi \right )^2 + e^{ 2 \alpha \phi} \cL \right )
\right ],
\label{a1}
\ee
where the Lagrangian $\cL$ is given by
\be
\cL = - {1 \over 2} \left (
\hna_{\alpha} \psi + i e A_{\alpha} \psi \right ) \hg^{\alpha \beta}
\left (
\hna_{\beta} \zpsi - i e A_{\beta} \zpsi \right ) - F_{\mu \nu}
F^{\mu \nu} .
\ee
The action is written in the {\it string frame} but it will be useful to rewrite it 
in the {\it Einstein frame}.
In {\it Einstein frame} the metric is related to the {\it string frame}
via the conformal transformation of the form provided by the following:
\be
g_{\alpha \beta} = e^{- 2 \phi} \hg_{\alpha \beta}.
\ee
The gravitational part of the action (\ref{a1}) appears in the {\it Einstein frame}
in a more familiar form. Namely, it implies the following:
\be
I = \int d^{4} x \sqrt{-g} \bigg[
R - 2 ( \na \phi )^{2} + e^{2 \alpha \phi + 4 \phi} \cL
(\psi, \zpsi, A, e^{2 \phi} g_{\alpha \beta}) \bigg].
\ee
The equations of motion derived from the variational principle yield
\ben
\na^{2} \phi &-&
{\alpha + 1 \over 4}~e^{2 \phi (\alpha + 1)}  \bigg( 
\na_{\beta} \psi + i e A_{\beta} \psi \bigg) 
\bigg( \na^{\beta} \zpsi - i e A^{\beta} \zpsi \bigg)
- {1 \over 2} \alpha
e^{ 2 \alpha \phi} F^{2} = 0,  \label{aaa} \\
 \na_{\mu} \bigg( e^{ 2 \alpha  \phi} F^{\mu \nu} \bigg) 
&+& {e^{2 \phi (\alpha + 1)}\over 4} \bigg[
i e \zpsi \bigg( \na^{\nu} \psi + i e A^{\nu} \psi \bigg)
- i e \psi \bigg( \na^{\nu} \zpsi - i e A^{\nu} \zpsi \bigg)
\bigg] = 0, \label{bbb}\\
\na^{2} \psi &+& i e A^{\beta} \bigg(
2 \na_{\beta} \psi + i e A_{\beta} \psi \bigg)
+ i e \na_{\delta} A^{\delta} \psi = 0, \label{ccc} \\
\na^{2} \zpsi &-& i e A^{\beta} \bigg(
2 \na_{\beta} \zpsi - i e A_{\beta} \zpsi \bigg)
- i e \na_{\delta} A^{\delta} \zpsi = 0, \label{c1c1c1} \\
G_{\mu \nu} &=& T_{\mu \nu}(\phi, F, \psi, \zpsi, A),
\een
where the energy momentum tensor $T_{\mu \nu}(\phi, F, \psi, \zpsi, A)$ for the fields in the theory
under consideration is provided by the relation
\be
T_{\mu \nu}(\phi, F, \psi, \zpsi, A) = e^{2 \phi( \alpha + 1)}
\tT_{\mu \nu}(\psi, \zpsi, A) + T_{\mu \nu}(F, \phi).
\label{ten}
\ee
In Eq.(\ref{ten}) by $\tT_{\mu \nu}(\psi, \zpsi, A)$ we have denoted the following expression:
\ben
\tT_{\mu \nu}(\psi, \zpsi, A) &=& 
{1 \over 4} \bigg[
ie ~\psi \bigg( 
A_{\mu}~\na_{\nu} \zpsi + A_{\nu}~\na_{\mu} \zpsi \bigg) - 
ie~\zpsi \bigg( 
A_{\mu}~\na_{\nu}\psi + A_{\nu}~\na_{\mu}\psi \bigg) \bigg]
\\ \nonumber
&+&
{1 \over 4} \bigg(
\na_{\mu} \psi~ \na_{\nu} \zpsi + \na_{\mu} \zpsi~ \na_{\nu} \psi \bigg) +
{1 \over 2}~e^2 A_{\mu}~ A_{\nu} \psi~ \zpsi +
{1 \over 2}~\tilde {\cal L}(\psi, \zpsi, A) g_{\mu \nu},
\een
where the explicit form of the Lagrangian $\tilde {\cal L}(\psi, \zpsi, A)$ is written as
\be
\tilde {\cal L}(\psi, \zpsi, A) = - {1 \over 2} \bigg(
\na_{\beta} \psi + i e A_{\beta} \psi \bigg)~
\bigg( \na^{\beta} \zpsi - i e A^{\beta} \zpsi \bigg).
\ee
On the other hand, for $T_{\mu \nu}(F, \phi)$ one gets
\be
T_{\mu \nu}(F, \phi) = e^{2 \alpha \phi} \bigg(
2 F_{\mu \rho} F_{\nu}{}{}^{\rho} - {1 \over 2} g_{\mu \nu} F^2 \bigg )
- g_{\mu \nu} (\na \phi)^2 + 2 \na_{\mu} \phi \na_{\nu} \phi.
\ee
In order to study the gravitational collapse in a spherically symmetric spacetime
it will be useful to consider the line element written in the
double null form \cite{chr}
\be
ds^2 = - a(u, v)^2 du dv + r^2(u, v) d \Omega^2,
\label{m}
\ee
where $u, v$ are advanced and retarded time null coordinates.
The null character of the coordinates in question will be preserved by the gauge transformation of the form
$u \rightarrow f(u)$ and $v \rightarrow g(v)$. Using doubly null coordinates enables us to begin with the regular
initial spacetime at approximately past null infinity, compute the
formation of black hole's event horizon and then prolong the evolution
of the black hole to the central singularity formed during the
dynamical collapse.\\
The assumption of spherical symmetry and the above coordinate choice imply that the only non-vanishing
component of the $U(1)$-gauge strength field is $F_{u v}$ or $F_{vu}$. Consequently, it provides another
restriction on the gauge potential. Namely, one has to do with $A_{u}$ or $A_{v}$. We can get rid of
one of these components of gauge potential by using the gauge freedom of the form
$A_{u} \rightarrow A_{u} + \na_{u} \theta$. If one chooses $\theta = \int A_{v}dv$, then we are left with the only one 
component of the gauge field, which is the  function of $u$ and $v$-coordinates. 
\par
To proceed further, we take into account the $v$-component of the generalized Einstein-Maxwell equations.
It leads to the following relation:
\be
\bigg[ {2 e^{2 \alpha \phi}~r^2~A_{u,v} \over a^2} \bigg]_{,v} +
{r^2~e^{2 \phi (\alpha + 1)} \over 4}~ie~  \bigg(
\zpsi~\psi_{,v} - \psi~\zpsi_{,v} \bigg) = 0.
\ee
Let us define the quantity
\be
Q = 2 {A_{u, v}~r^2 \over a^2},
\label{charge}
\ee 
just as in Ref.\cite{ore03}. $Q$ corresponds to the electric charge within the sphere of the radius
$r(u, v)$.
The above definition enables us to separate the second order partial differential 
equation for $A_{u}$ into two much simpler first order differential equations.
We arrive at the following:
\be
A_{u, v} - {Q a^{2} \over 2 r^{2}} = 0,
\label{p}
\ee
and
\be
Q_{,v} + 2~ \alpha~ \phi_{,v}~ Q + {i e~ r^2 \over 4} e^{2 \phi}
\bigg( \zpsi \psi_{,v} - \psi \zpsi_{,v} \bigg) = 0. 
\label{l}
\ee
The equation of motion for dilaton field (\ref{aaa}) has the form provided by
\be
r_{,u} \phi_{,v} + r_{,v} \phi_{,u} + r \phi_{,uv} -
{(\alpha + 1) \over 8}~e^{2 \phi (\alpha + 1)}~ r~ \bigg[
\psi_{,u} \zpsi_{,v} + \psi_{,v} \zpsi_{,u} + i e~ A_{u}
\bigg( \psi \zpsi_{,v} - \zpsi \psi_{,v} \bigg ) \bigg ]
- \alpha
{a^2~Q^2~ e^{ 2 \alpha \phi}
\over 4 r^3} = 0.
\ee
Consequently, the relations for the complex scalar fields are given by
\ben
r_{,u} \psi_{,v} + r_{,v} \psi_{,u} + r \psi_{,uv} + i e~ r~ A_{u}~ \psi_{,v}
+ i e~ r_{,v}~ A_{u}~ \psi + {i e~ Q~ a^2 \over 4 r} \psi = 0, \\
r_{,u} \zpsi_{,v} + r_{,v} \zpsi_{,u} + r \zpsi_{,uv} - i e~ r~ A_{u}~ \zpsi_{,v}
- i e~ r_{,v}~ A_{u}~ \zpsi - {i e~ Q~ a^2 \over 4 r} \zpsi = 0.
\een
Combining the adequate components of the Einstein tensor and the stress-energy tensor for the underlying
theory we obtain the following set of equations:
\ben
{2 a_{,u}~ r_{,u} \over a} - r_{,u u}
&=& r~ \phi_{,u}^2 +
{r~e^{2 \phi (\alpha + 1)} \over 4} \bigg[
\psi_{,u} \zpsi_{,u} + i e~ A_{u}~ \bigg(
\psi~\zpsi_{,u} - \zpsi~\psi_{,u} \bigg ) + e^2 ~A_{u}^2 \psi~ \zpsi 
\bigg], \\
{2 a_{,v}~ r_{,v} \over a} - r_{,vv} &=& 
 r~ {\phi_{v}}^2 + {1 \over 4}~r~ e^{2 \phi (\alpha + 1)}~ \psi_{,v}~ \zpsi_{,v}, \\
{a^2 \over 4r} + {r_{,u}~ r_{v} \over r} +  r_{,uv} &=& {e^{2 \alpha \phi}~a^2~Q^2 \over 4 r^3}, \\
{a_{,u} a_{,v} \over a^2}
- {a_{,uv} \over a} - {r_{,uv} \over r} &=&
{Q^2~ e^{2 \alpha \phi}~ a^2 \over 4 r^4} +  \phi_{,u} \phi_{,v} 
+ {e^{2 \phi(\alpha + 1)} \over 8} \bigg [
\psi_{,u}~ \zpsi_{,v} + \zpsi_{,u}~ \psi_{,v}
+ i e~ A_{u}~ \bigg (\psi~ \zpsi_{v} - \zpsi~\psi_{,v} \bigg) 
\bigg].
\een
Moreover, we introduce new auxiliary variables written in the form as
\ben
c = \frac{a_{,u}}{a}, \qquad d = \frac{a_{,v}}{a},
\qquad f = r_{,u}, \qquad g = r_{,v},  \nonumber\\
s = \psi,  \qquad  p  =  \psi_{,u}, \qquad q  =  \psi_{,v}, \qquad \beta  =  A_u,\\
k = \phi, \qquad x = \phi_{,u}, \qquad y  = \phi_{,v},  \nonumber
\label{eqn:substitution}
\een
and the additional quantities provided by the relations as follows:
\ben
\la & \equiv & \frac{a^2}{4} + f g,
\label{eqn:lambda-definition}\\
\mu & \equiv & f q + g p,
\label{eqn:mi-definition}\\
\eta & \equiv & g x + f y.
\label{eqn:eta-definition}
\een
Instead of considering two complex fields $\psi$ and $\zpsi$ one can introduce two real fields
obeying the relations
$\psi = \psi_{1} + i~\psi_{2}$ and $\zpsi = \psi_{1} - i~\psi_{2}$. 
On this account, it leads to                       
\ben \label{eqn:complex-functions-s-p-q}
s &=& s_1 + i~s_2, \qquad p  =  p_1 + i ~p_2, \qquad q  =  q_1 + i ~q_2, \\ \label{eqn:complex-function-mu}
\mu & = & \mu_1 + i ~\mu_2, \qquad
\mu_1 = f q_1 + g p_1, \qquad \mu_2  =  fq_2 + gp_2.
\label{eqn:complex-functions-mu1-mu2}
\een
Thus, having in mind all the above, one can rewrite the system of the second order partial differential 
equations as the first order one.
By this procedure we get the following system of the first order differential equations:
\ben \label{eqn:P1-2}
& P1: & a_{,u} - a~c  =  0,\\
& P2: & a_{,v}- a~d =  0,\\
& P3: & r_{,u}- f  =  0,\\
& P4: & r_{,v} - g  =  0,\\
& P5_{_{\left(Re\right)}}: & s_{1,u} - p_1  =  0,\\
& P5_{_{\left(Im\right)}}: & s_{2,u} - p_2  =  0,\\
& P6_{_{\left(Re\right)}}: & s_{1,v} - q_1  =  0,\\
& P6_{_{\left(Im\right)}}: & s_{2,v} - q_2  =  0,\\
& P7: & k_{,u} - x  =  0,\\
& P8: & k_{,v} - y  =  0,\\
& E1: & f_{,u} - 2~c~f + r~x^2+\frac{1}{4}~r~e^{2k\left(\alpha+1\right)}
\Big[p_1^{\: 2} + p_2^{\: 2} + 2~e~\beta~\left(s_1~p_2 - s_2~p_1 \right)
+ e^2~\beta^2~\left(s_1^{\: 2} + s_2^{\: 2} \right) \Big]  =  0,\\
& E2: & g_{,v} - 2~d~g +  r~y^2 + \frac{1}{4}~r~e^{2k\left(\alpha+1\right)}~\left(q_1^{\: 2} + q_2^{\: 2}\right)
= 0,\\
& E3^{\left(1\right)}: & f_{,v} + \frac{\la}{r} - e^{2\alpha k}~\frac{Q^2a^2}{4~r^3}  =  0,\\
& E3^{\left(2\right)}: & g_{,u} + \frac{\la}{r} -e^{2\alpha k}~\frac{Q^2a^2}{4~r^3}  =  0,\\
& E4^{\left(1\right)}: & c_{,v} - \frac{\la}{r^2} + x~y + \frac{1}{4}~e^{2k\left(\alpha+1\right)}
\Big[p_1~q_1 + p_2~q_2 + e~\beta~\left(s_1~q_2 - s_2~q_1 \right) \Big] + e^{2\alpha k} \frac{Q^2a^2}{2~r^4}  = 0,\\
& E4^{\left(2\right)}: & d_{,u} - \frac{\la}{r^2} + x~y
+ \frac{1}{4}~e^{2k\left(\alpha+1\right)}
\Big[p_1~q_1 + p_2~q_2 + e~\beta~\left( s_1~q_2 - s_2~q_1 \right) \Big] + e^{2\alpha k}\frac{Q^2a^2}{2~r^4}  =  0,\\
& S_{_{\left(Re\right)}}^{\left(1\right)}: & rp_{1,v} + \mu_1 - e~ r~\beta~ q_2 - e~ s_2~\beta~ g -
e~ s_2~\frac{Qa^2}{4r}  =  0,\\
& S_{_{\left(Im\right)}}^{\left(1\right)}: & rp_{2,v} + \mu_2 + e~ r~\beta~ q_1 + e~ s_1~\beta~ g + e~ s_1~\frac{Qa^2}{4r}
= 0,\\
& S_{_{\left(Re\right)}}^{\left(2\right)}: & rq_{1,u} + \mu_1 - e~ r~\beta~ q_2 - e~ s_2~\beta~ g - e~ s_2~\frac{Qa^2}{4r}
= 0,\\
& S_{_{\left(Im\right)}}^{\left(2\right)}: & rq_{2,u} + \mu_2 + e~ r~\beta~ q_1 + e~ s_1~\beta~ g + e~ s_1~\frac{Qa^2}{4r}
= 0,\\
& D^{\left(1\right)}: & rx_{,v} + \eta - \frac{\alpha+1}{4}~r~e^{2k\left(\alpha+1\right)}
\Big[p_1~q_1 + p_2~q_2 + e~\beta~\left(s_1~q_2 - s_2~q_1 \right) \Big] - \alpha~ e^{2\alpha k}~\frac{Q^2a^2}{4~r^3}  =  0,\\
& D^{\left(2\right)}: & r~y_{,u} + \eta - \frac{\alpha+1}{4}~r~e^{2k\left(\alpha+1\right)}
\Big[p_1~q_1 + p_2~q_2 + e~\beta~\left( s_1~q_2 - s_2~q_1 \right) \Big] - \alpha~ e^{2\alpha k}\frac{Q^2a^2}{4~r^3}  =  0,\\
& M1: & \beta_{,v} - \frac{Qa^2}{2~r^2}  =  0,\\
& M2: & Q_{,v} + 2~\alpha~ y~Q - {1 \over 2}~e^{2k}~e~ r^2 \left( s_1~q_2 - s_2 ~q_1 \right)  =  0.
\label{eqn:M2-2}
\een
Let us introduce some quantities of physical interest.
Namely, we define the mass function provided by the relation
\be
m(u,v) = {r \over 2}~\bigg( 1 + {4 ~r_{,u}~r_{,v} \over a^2} \bigg) =
{r \over 2}~\bigg( 1 + {4 \over a^2}~f~g \bigg) .
\label{mhaw}
\ee
It represents the Hawking mass, i.e., the mass included in a sphere of the radius $r(u,v)$.
Moreover, the Ricci scalar has the form as
\be
R\left(u,v\right)  =  -\frac{16 x
~y}{a^2} - \frac{2}{a^2}~e^{2k\left(\alpha+1\right)}~\Big[p~q^\ast + q ~p^\ast +
i~e~\beta~\left( s~q^\ast -q ~s^\ast \right) \Big],
\label{eqn:Ricci-scalar-u-v-1}
\ee
or it can be rewritten in the form which yields the following:
\be
R\left(u,v\right)  =  -\frac{16 x~y}{a^2} - \frac{4}{a^2}~ e^{2k\left(\alpha+1\right)}~\Big[p_1~q_1 + p_2~q_2 +
e~\beta~\left( s_1~q_2 - s_2~q_1 \right) \Big].
\label{eqn:Ricci-scalar-u-v-2}
\ee


\section{Numerical computations}
\subsection{Numerical algorithm}
The system of equations (\ref{eqn:P1-2})-(\ref{eqn:M2-2}) in the theory under consideration
has to be solved numerically. In order to find the solution one should elaborate
an evolution of the quantities $d$, $q_1$, $q_2$, $y$, $a$, $s_1$, $s_2$, $k$, 
$g$, $r$, $Q$, $\beta$, $f$, $p_1$, $p_2$ and $x$. The quantity $c$ does not play the significant role in the process
under consideration, so it can be ignored. 
The evolution of the quantities
$d$, $q_1$, $q_2$ and $y$ along $u$ is governed by relations $E4^{\left(2\right)}$, 
$S_{\left(Re\right)}^{\left(2\right)}$, $S_{\left(Im\right)}^{\left(2\right)}$ and $D^{\left(2\right)}$, respectively.
The remaining quantities, $a$, $s_1$, $s_2$, $k$, $g$, $r$, $Q$, $\beta$, $f$, $p_1$, $p_2$ and $x$
evolve in turn along $v$-coordinate
according to equations $P2$, $P6_{\left(Re\right)}$, $P6_{\left(Im\right)}$, $P8$, $E2$, $P4$, $M2$, $M1$, 
$E3^{\left(1\right)}$, $S_{\left(Re\right)}^{\left(1\right)}$, $S_{\left(Im\right)}^{\left(1\right)}$ and $D^{\left(1\right)}$. 
On the other hand, Eqs. $P1$ and $E4^{\left(1\right)}$ describing the behaviour of $c$ may 
be discarded and the remaining relations can be used to determine the boundary conditions.
\par
In our studies the numerical algorithm similar to the one proposed in \cite{ham96} was implemented. 
The computations were carried out on the two-dimensional grid constructed in the $(vu)$-plane. 
In order to obtain a value of a particular function at a point $\left(v,u\right)$ 
one should have values of the appropriate functions at points $\left(v-h_v,u\right)$ and $\left(v,u-h_u\right)$, 
where $h_v$ and $h_u$ are integration steps in $v$ and $u$ directions, respectively. 
Equations describing the evolution of the considered quantities along 
the coordinates $u$ and $v$ may be symbolically written as:
\ben
\textsf{f}_{,u} \ \; = \ \; \textsf{F}\left(\textsf{f},\textsf{g}\right), \qquad
\textsf{g}_{,v} \ \; = \ \; \textsf{G}\left(\textsf{f},\textsf{g}\right).
\een
In order to get the value of the particular function at a point $\left( v,u \right)$,
we should find the auxiliary quantities which yield
\ben
\textsf{ff}\big\arrowvert_{\left(v,u\right)} 
& = & \textsf{f}\big\arrowvert_{\left(v,u-h_u\right)} + 
h_u\textsf{F}\left(\textsf{f},\textsf{g}\right)\big\arrowvert_{\left(v,u-h_u\right)},
\label{eqn:auxiliary-quantity-u}\\ \nonumber \\
\textsf{gg}\big\arrowvert_{\left(v,u\right)} 
& = & \textsf{g}\big\arrowvert_{\left(v-h_v,u\right)} +
\frac{h_v}{2}\bigg(\textsf{G}\left(\textsf{f},\textsf{g}\right)\big\arrowvert_{\left(v,u\right)} +
\textsf{G}\left(\textsf{ff},\textsf{gg}\right)\big\arrowvert_{\left(v,u\right)}\bigg).
\label{eqn:auxiliary-quantity-v}
\een
By virtue of the above the final values 
of the quantities in question are provided by the following:
\ben
\textsf{f}\big\arrowvert_{\left(v,u\right)} 
& = & \frac{1}{2}\bigg(\textsf{ff}\big\arrowvert_{\left(v,u\right)}+
\textsf{f}\big\arrowvert_{\left(v,u-h_u\right)} +
h_u\textsf{F}\left(\textsf{ff},\textsf{gg}\right)\big\arrowvert_{\left(v,u\right)}\bigg),\\ \nonumber \\
\textsf{g}\big\arrowvert_{\left(v,u\right)} 
& = & \frac{1}{2}\bigg(\textsf{gg}\big\arrowvert_{\left(v,u\right)}+
\textsf{g}\big\arrowvert_{\left(v-h_v,u\right)} +
h_v\textsf{G}\left(\textsf{ff},\textsf{gg}\right)\big\arrowvert_{\left(v,u\right)}\bigg).
\een
In the early stages of the calculations the numerical grid is divided evenly, both in $v$ and $u$ directions.
On this account, at the beginning the quantities $h_v$ and $h_u$ are equal to each other.

\subsection{Adaptive mesh refinement}
The coordinates $u$ and $v$ ensure the regular behaviour of all the considered quantities within 
the domain of integration except the vicinity of $r=0$. However, during the numerical analysis 
the considerable difficulties also arise close to the event horizon, where function $f$ 
diverges. A relatively dense numerical grid is necessary in order to satisfactorily 
determine the location of the event horizon and to examine the behaviour of fields 
inside it, especially for large values of the $v$-coordinate. 
The efficiency of the calculations suggests using
an adaptive grid and performing integration with a smaller step in particular 
regions. 
On this account the refinement algorithms 
enable us to make the grid denser both in $v$ and $u$ directions and to do the same
only along $u$-coordinate. 
The first one makes the integration steps $h_v$ and $h_u$ smaller 
on the equal footing
as one approaches the event horizon and reaches the large values of $v$. 
The other one changes only the value of $h_u$. It turned out that the latter 
refinement of the adaptive grid gives the satisfactory results and it is more effective
due to the computation time and required computer's memory.
Hence, all the results presented in our paper will be based on it.
\par
In order to determine the area of the integration grid, where 
the grid should be denser,
a local error indicator need to be used. 
The aforementioned quantity ought to be bounded with the
evolving quantities as well as it should change its value significantly in the adequate region.
It happened that \cite{ore03} the function ${\Delta r}/r$ along $u$-coordinate
meets our requirements.

\subsection{Boundary and initial conditions}
Having specified the numerical algorithm for the solution of the equations of motion we refine
our studies to the case of the initial and boundary conditions for the equations in question.
The boundary conditions refer to the
surface $u=v$, while the initial conditions are formulated along an arbitrarily chosen constant surface $u = u_i$. 
A point $\left(0,0\right)$  is chosen to be the intersection of these two lines in the $(vu)$-plane.
It also indicates the center of the considered spacetime.
\par
The physical situation we shall take into account will be the gravitational collapse of a spherically symmetric
shell of infalling complex charged matter. Spacetime of a spherical shell of matter
is flat in two regions, i.e., inside the shell and at large radii from it. This fact enables us to
assume that the line $u=v$ will be not significantly affected by the presence of the collapsing shell
of matter.
On this account the spacetime may be considered as nearly 
flat and electrically neutral in that region. 
Thus, it gives the following boundary conditions: $r = Q = \beta = 0$. 
Consequently, equations $E3^{\left(1,2\right)}$ reveal that $\la=0$ along $u=v$. This fact
together with Eqs.$P3$ and $P4$ provides that $f = - g = - \frac{a}{2}$. 
Furthermore, the requirement $s_{1,r}=s_{2,r}=k_{,r}=a_{,r}=0$ 
along $u=v$ guarantees that the field functions flatten near $r=0$ making the numerical analysis possible. 
Because of the fact that $r$ changes non-linearly along the coordinates $u$ and $v$ this boundary condition is 
implemented using the three-point regressive derivative method with a variable step,
except the first point, where the Euler's method is used. 
\par
Combining the relation $\mu_1 = \mu_2 = \eta = 0$ along $u=v$ line and 
Eqs.$S_{\left(Re\right)}^{\left(1,2\right)}$, we obtain that $p_1=q_1$, $p_2=q_2$ and $x=y$.
On the other hand, the boundary conditions for the quantities evolving 
along $u$ can be achieved according to the algorithm described in the previous section.
It can be seen that the auxiliary quantities (\ref{eqn:auxiliary-quantity-v})
are taken to be equal to the corresponding functions at $r=0$.
\par
The assumption of the flat geometry in the region, where an observation of the collapse begins justifies 
the condition that $d\left(v,0\right) = 0$. Further, it fixes the remaining freedom in $v$-coordinate. 
By virtue of the flatness of the spacetime 
in the vicinity of a surface $u=v$ and the above assumptions we get that $a\left(v,0\right)=1$.  
One should mention that the initial conditions ought to include the arbitrary profiles for scalar
functions $s_1\left(v,0\right)$ and $s_2\left(v,0\right)$ describing the
real and the imaginary parts of the charged scalar field and for the dilaton field $k\left(v,0\right)$.
The one-parameter families of the initial profiles are listed in Table \ref{tab:initial-profiles},
where a free family parameter is denoted as $\tpe$, constants $c_1$ and $c_2$ are arbitrarily chosen, 
while $v_f$ equals the final value of $v$. $\delta\in\left[0,\frac{\pi}{2}\right]$ 
is a phase difference determining the amount of the initial electric charge \cite{hwa10a}.
The quantities $q_1\left(v,0\right)$, $q_2\left(v,0\right)$ 
and $y\left(v,0\right)$ are computed analytically using equations $P6_{\left(Re\right)}$, $P6_{\left(Im\right)}$ and $P8$. 
The values of the functions $g$, $r$, $Q$, $\beta$, $f$, $p_1$, $p_2$ and $x$ along the axis $u=0$ are 
obtained using the three-point Simpson's method apart from the first point, where the Newton's method is implemented.

\begin{table}
\caption{Initial profiles of field functions}
\begin{ruledtabular}
\begin{tabular}{c|cp{5cm}|}
Family & Profile\\
\hline
&\\
$\left(f_{D+S}\right)$ & $\tpe\cdot v^2\cdot e^{-\left(\frac{v-c_1}{c_2}\right)^2}$\\
&\\
$\left(f_S\right)$ & $\tpe\cdot \sin^2\left(\pi\frac{v}{v_f}\right)
\cdot\Bigg(\cos\left(\pi\frac{2v}{v_f}\right)+i\cos\left(\pi\frac{2v}{v_f}+\delta\right)\Bigg)$\\
&\\
\end{tabular}
\end{ruledtabular}
\label{tab:initial-profiles}
\end{table}
\subsection{Numerical tests}
The most straightforward manner of checking the correctness of the numerical code will be a comparison between 
achieved results and an analytical solution of the considered problem. Unfortunately,
because of the lack of the analytical solution of the problem in question one should apply indirect methods of
checking the accuracy of the numerical code.
\par
\begin{figure}[p]
\includegraphics[scale=0.75]{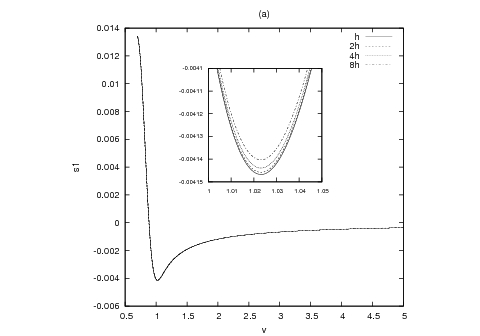}\\
\includegraphics[scale=0.75]{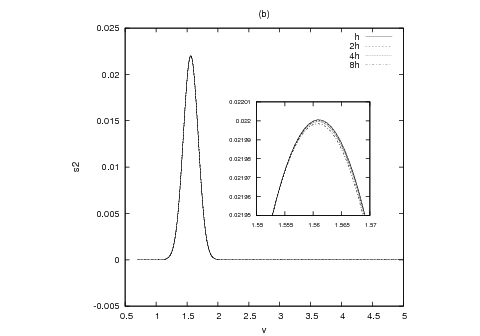}\\
\includegraphics[scale=0.75]{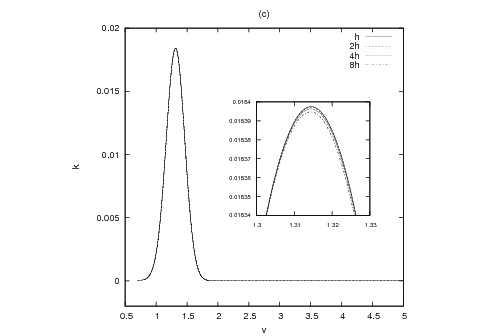}
\caption{Plots of the field profiles: (a) real and (b) imaginary parts of electrically charged scalar field,
(c) dilaton field, along $u=0.7$ taken for the different integration steps 
(multiples of $h=10^{-4}$). The evolution was monitored for a Gaussian initial 
pulse $\left(f_{D+S}\right)$ with the parameter $\tpe=0.005$ and constants $\left(c_1,c_2\right)$ equal to 
$\left(0.75,0.15\right)$, $\left(1.55,0.17\right)$, $\left(1.3,0.21\right)$, respectively. The values of 
coupling constants are $\alpha=1$ and $e = 0.5$.}
\label{fig1}
\end{figure}
\begin{figure}[p]
\includegraphics[scale=0.75]{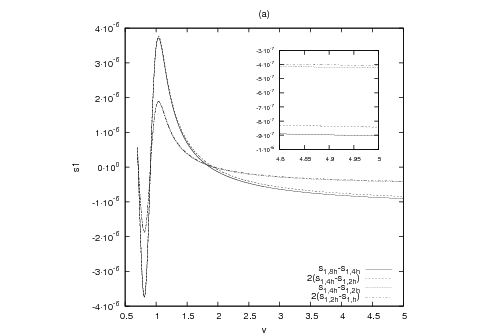}\\
\includegraphics[scale=0.75]{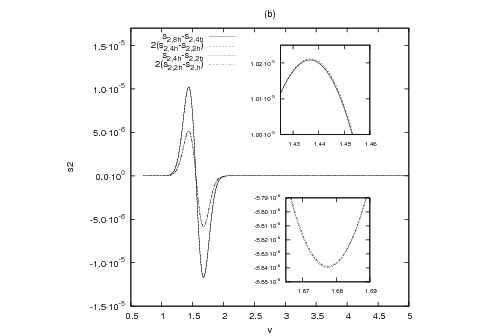}\\
\includegraphics[scale=0.75]{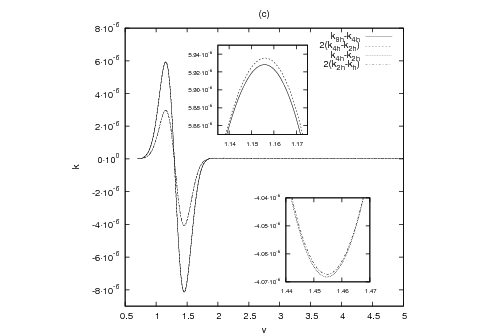}
\caption{The differences between the field profiles: (a) real and (b) imaginary parts of electrically 
charged scalar field, (c) dilaton field and their corresponding doubles for
evolutions elaborated in Fig.\ref{fig1}. The subscripts of field functions denote integration steps (multiples $h=10^{-4}$).}
\label{fig2}
\end{figure}

The first trial will be checking of the convergence of the obtained results.
One verifies that the fields under consideration converge to some values in the expected region of
convergence. From the numerical point of view it envisages the fact that the algorithm and its
implementation are free of mistakes. 
To begin with we carried the computations not requiring adaptive mesh on four different grids 
with integration steps equal in both directions. The integration step of the particular grid 
was twice the size of a denser one. The evolving field profiles for 
arbitrarily chosen $u$-coordinate are shown in Fig.\ref{fig1}. 
We scaled up the vicinities of the cusps, where the differences 
among profiles were most significant. 
For all the field profiles the very good agreement of 
an order of $0.01\%$ was achieved. 
On the other hand, the linear convergence of the numerical code is presented in
Fig.\ref{fig2}. 
The differences between  
the profiles obtained on the two grids with a quotient of the integration steps equal to $2$ and their
respective
doubles are  hardly distinguishable. The divergence is at most $1\%$, as may be inferred from the 
values shown in the magnified areas. Moreover, Fig.\ref{fig2} gives 
the clear evidence that the errors become smaller as the grid density increases.
\par
\begin{figure}[p]
\centering
\includegraphics[scale=1.25]{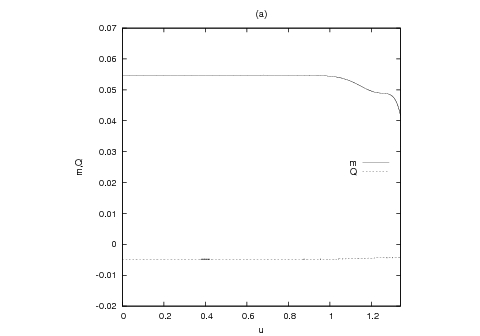}
\includegraphics[scale=1.25]{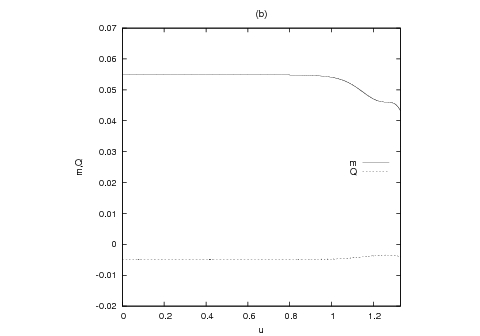}
\caption{Mass and charge as functions of $u$-coordinate along $v=5$ outside event horizon for: 
(a) $\alpha=-1$ and (b) $\alpha=1$. Initial profile of the dilaton field was a Gaussian pulse 
$\left(f_{D+S}\right)$ with constants $\left(c_1,c_2\right)$ equal to $\left(0.17,0.21\right)$. 
The charged scalar field was initially represented by the
profile $\left(f_S\right)$ with constants $v_f=5$ and $\delta=\frac{\pi}{2}$. The value of 
electric coupling constant is $e = 0.5$, the free parameter is taken to be $\tpe=0.075$.}
\label{fig3}
\end{figure}

The next test of our code is to check whether mass (\ref{mhaw}) and
charge (\ref{charge}) are conserved in the evolving spacetime. 
The considered fields are scattered by the gravitational and electromagnetic potential barriers as 
the collapsing shell approaches its gravitational radius. Therefore
the conservation laws are not satisfied in the entire domain of integration. Nevertheless,
this effect of the outgoing fluxes of mass and charge is negligible \cite{ore03} and it has no
significant influence on the total mass and charge, excluding the area in the vicinity
of new forming black hole event horizon. In Fig.\ref{fig3} the behaviours of mass and charge for the large
value of the advanced time are presented. It turned out that for $u$ not exceeding $1$,
mass and charge are conserved up to within $1.6\%$ and $2.5\%$, respectively.
Further, the inspection of Fig.\ref{fig3} reveals the deviation of 
the aforementioned quantities from constancy increases with the advanced time. It is caused due to the fact
that the reflected waves carry off some mass as well as charge. Because of the fact that the total mass
contains also the energy momentum of gravitational field which is not taken into account in the Noether
current bounded with the energy momentum tensor, the subject of the mass conservation is more subtle \cite{ore03}.
\par
\begin{figure}[p]
\centering
\includegraphics[scale=1.25]{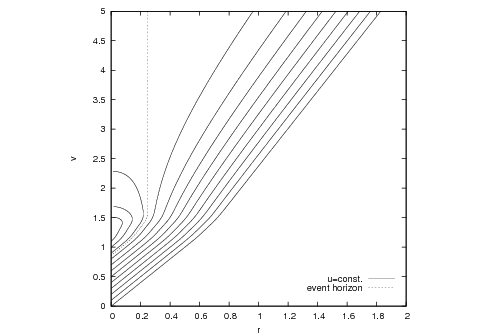}
\caption{Outgoing null rays in $(rv)$-plane for the spacetime containing a black hole left 
after gravitational collapse of a neutral scalar field. The initial profile belongs to the 
family $\left(f_{D+S}\right)$ with constants $\left(c_1,c_2\right)$ equal to $\left(1.3,0.21\right)$. 
The value of the free parameter is $\tpe=0.125$.}
\label{fig4}
\end{figure}
\begin{figure}[p]
\centering
\includegraphics[scale=1.25]{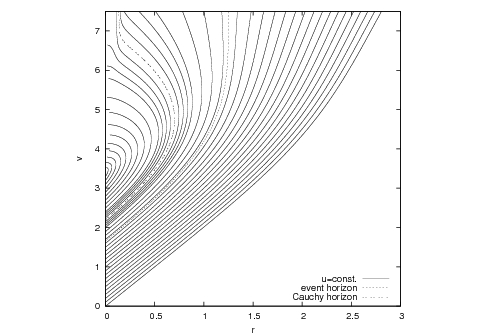}
\caption{Outgoing null rays in $(rv)$-plane for the spacetime containing a black hole 
left after gravitational collapse of an electrically charged scalar field. The initial profile belongs 
to the family $\left(f_S\right)$ with constants equal to $v_f=7.5$ and $\delta=\frac{\pi}{2}$. 
The value of electric coupling constant is $e=0.5$ and the value of free parameter is $\tpe=0.5$.}
\label{fig5}
\end{figure}

The last test of the accuracy of our code consists of the analysis of the simplified
versions of the problem in question.
Namely, in Fig.\ref{fig4} we depicted the outgoing null rays in $(rv)$-plane for the spacetime containing
black hole stemmed from the gravitational collapse of the neutral scalar field. 
The situation corresponds to setting $\alpha = e = 0$ in the equations of motion
and eliminating electrically charged scalar field by putting $\psi = 0$.
The value of the free
parameter is equal to $0.125$. On the other hand, Fig.\ref{fig5} illustrates the outgoing null rays
for the spacetime of a black hole emerging due to the gravitational collapse of an electrically charged scalar fields.
Here, we put $\alpha = 0$ and get rid of the dilaton field by setting $\phi = 0$. The value of free parameter was taken to be
$0.5$.
The inspection of Fig.\ref{fig5} shows the formation of the black hole event horizon and a Cauchy horizon at asymptotically
large $v$.
The structures of the spacetimes emerging during collapses presented in Figs.\ref{fig4} and \ref{fig5} are in a perfect
agreement with the results published in Refs.\cite{ham96,hod98,ore03}.


\section{Results}
In our numerical studies we have used the
one-parameter families of initial profiles referring to the real and 
imaginary parts of the electrically charged scalar field and dilaton field. 
The results do not depend on the type of the family of the initial profiles as well as on family constants. Hence
their choice is unrestricted. Moreover, no significant dependence on the electric coupling constant was observed.
On this account the electric coupling constant was put $e = 0.5$, in all the calculations.
All 
the results in the present section were obtained using profiles $\left(f_{D+S}\right)$ with values of the family 
constants equal respectively to: $c_{1,s_1}=0.75$, $c_{2,s_1}=0.15$, $c_{1,s_2}=1.55$, $c_{2,s_2}=0.17$ and 
$c_{1,k}=1.3$, $c_{2,k}=0.21$. 
Subscripts $s_{1},~s_{2}$ and $k$ refer to the real and imaginary parts of the 
electrically charged fields and dilaton field, respectively.
The
family parameters $\tpe_{s_1}$, $\tpe_{s_2}$ and $\tpe_{k}$ 
are taken to be equal and they are denoted by $\tpe$.

The critical phenomena in gravitational physics turned out to be one of the key thought
experiments in the studies of black hole formation (see Ref.\cite{cho93,gun03} and references therein).
There is a critical value of the parameter, denoted by $\tpe^\ast$, below which the spacetime is 
non-singular and does not contain a black hole. For values exceeding $\tpe^\ast$ there 
is a black hole in the spacetime, which means that it is singular. These phenomena
are referred to as subcritical and supercritical, respectively. 
It was revealed in Ref.\cite{fro03} that for any fixed radius observer, as we take the limit 
to the critical value of the parameter,
the sphere of the influence  of the diminishing black hole mass shrinked to zero. On the other hand, the resulting
spacetime converges pointwise to Minkowski spacetime at $r > 0$. Moreover, the convergence is not even
uniform.
\par
\begin{figure}[p]
\includegraphics[scale=1.25]{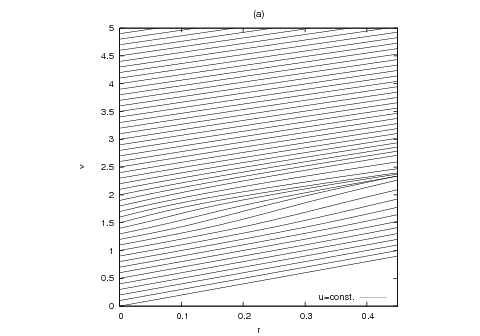}
\includegraphics[scale=1.25]{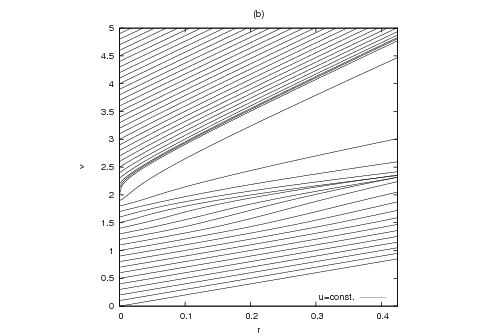}
\caption{Outgoing null rays in $(rv)$-plane for: (a) slightly curved ($\tpe\ll \tpe^\ast$) and 
(b) almost critical ($\tpe\lesssim \tpe^\ast$) non-singular spacetime left after the evolution of the discussed fields. 
The values of the free parameters are $\tpe=0.05$ and $\tpe=0.0517693$, respectively.
The coupling constants are $\alpha=1$ and $e=0.5$.}
\label{fig6}
\end{figure}

In what follows we shall study the gravitational collapse of a self-interacting complex charged scalar 
field in the dilaton gravity. We take into account models with different values of coupling constant $\alpha$.
To begin with one considers first a subcritical evolution.
In Fig.\ref{fig6} and \ref{fig8} we depicted
the radial function $r(u, v)$ as a function of the ingoing null coordinate $v$ along a sequence of the
outgoing null rays $(u = const.)$ All the outgoing null rays originate from the nonsingular axis $r = 0$.
Just
we begin our evolution with a regular spacetime
along $u = 0$. In Figs.\ref{fig6} and \ref{fig8} we plotted two cases, first one for a 
slightly curved spacetime when $\tpe\ll \tpe^\ast$ and the other for almost critical one, for which 
$\tpe\lesssim \tpe^\ast$. We take into account two values of free parameter $\tpe = 0.05$ and $\tpe=0.0517693$,
for the case when the coupling constant $\alpha = 1$ and $e = 0.5$, and
$\tpe = 0.0485$ and $\tpe=0.0503946$, when the value of coupling constant $\alpha =- 1$ and $e = 0.5$.
In both figures we observe a flat region corresponding to small values of advanced and retarded times in 
which one has that $r(u, v)$ is proportional to $u$ and $v$. With the passage of time
the curvature of the spacetime becomes considerable and we have no longer proportionality
between $r$ and $(u, v)$. As was expected the spacetime under consideration curved
earlier and more significantly for larger value of the parameter $\tpe$.
In Fig.\ref{fig6} we observe the intersections of $u = const.$ lines for larger values of $r$.
\par
\begin{figure}[p]
\includegraphics[scale=1.25]{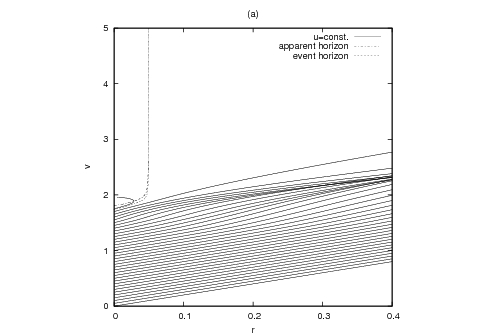}
\includegraphics[scale=1.25]{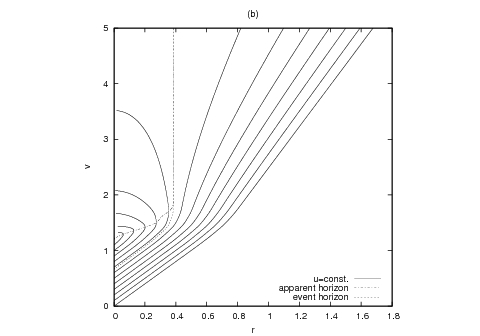}
\caption{Outgoing null rays in $(rv)$-plane for spacetime containing: 
(a) a small black hole ($\tpe\gtrsim \tpe^\ast$) and (b) a large black hole ($\tpe\gg \tpe^\ast$) 
emerged from the gravitational collapse under consideration. 
Formation of the apparent and event horizons is depicted. The values of the free parameters are $\tpe=0.0525$ and $\tpe=0.1$, respectively.
The coupling constants $\alpha$ and $e$ are the same as in Fig.\ref{fig6}.}
\label{fig7}
\end{figure}
\begin{figure}[p]
\includegraphics[scale=1.25]{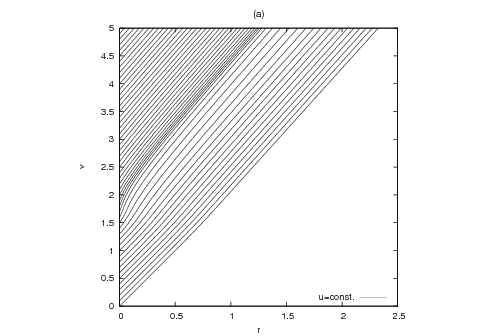}
\includegraphics[scale=1.25]{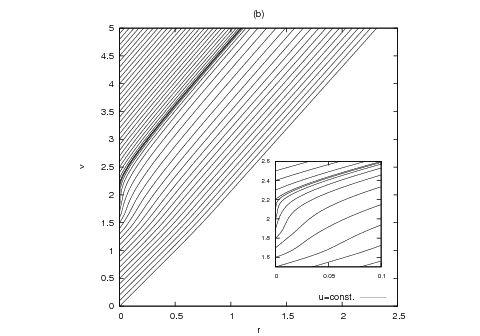}
\caption{Outgoing null rays in $(rv)$-plane for: (a) slightly curved and 
(b) almost critical non-singular spacetime left after the evolution of the fields in question. 
The free parameters are taken to be $\tpe=0.0485$ and $\tpe=0.0503946$, respectively. 
The coupling constants are $\alpha=-1$ and $e=0.5$.}
\label{fig8}
\end{figure}
\begin{figure}[p]
\includegraphics[scale=1.25]{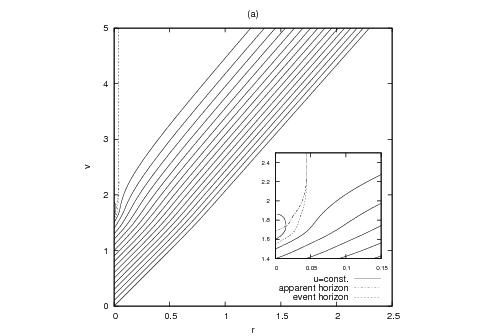}
\includegraphics[scale=1.25]{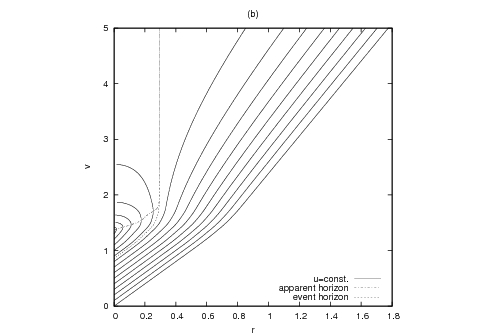}
\caption{Outgoing null rays in $(rv)$-plane for spacetime containing: 
(a) a small and (b) a large black hole emerged from the gravitational collapse under consideration. 
Formation of an apparent and an event horizons is depicted. 
The values of the free parameters are: $\tpe=0.0525$ and $\tpe=0.1$. The values of 
the coupling constants are the same as in Fig.\ref{fig8}.}
\label{fig9}
\end{figure}

Figs.\ref{fig7} and \ref{fig9} display Kruskal diagrams $(r(u, v), v)$ for the spacetimes under consideration.
As in the previous case we start the evolution from a regular spacetime
along $u = 0$. We studied two cases of singular spacetime, one with small black hole
when $\tpe\gtrsim \tpe^\ast$ and the other one, with big black hole for which $\tpe\gg \tpe^\ast$. The values of the $\tpe$
parameter are $0.0525$ and $0.1$, respectively. Fig.\ref{fig7} describes the birth of black holes for the coupling constant
$\alpha = 1$ and $e = 0.5$, while Fig.\ref{fig9} is performed for the case $\alpha = - 1$ and $e = 0.5$.
In both figures one can distinguish between two types of outgoing null rays in the $(rv)$-plane. Namely,
the outermost null rays escaping to the future null infinity (for small value of retarded time) and the innermost
null rays which stem from the nonsingular axis $r = 0$ and terminate at the singular part of the hypersurface
$r = 0$. Their evolution is described by finite value of $v$. Contrary to Refs.\cite{hod98,ore03}
we did not find in our numerical simulations intermediate outgoing null rays approaching
a fixed radius at late times (when $v \rightarrow \infty$). 
Just, the evolution in question resembles
formation of Schwarzschild black hole, rather than RN one with Cauchy horizon.
\par
\begin{figure}[p]
\includegraphics[scale=1.25]{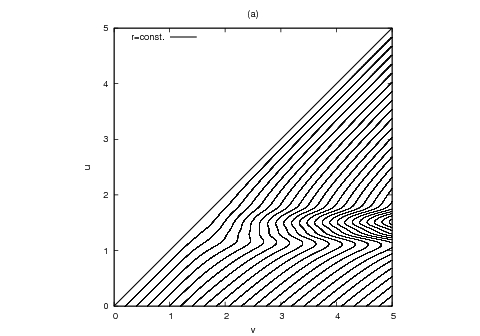}
\includegraphics[scale=1.25]{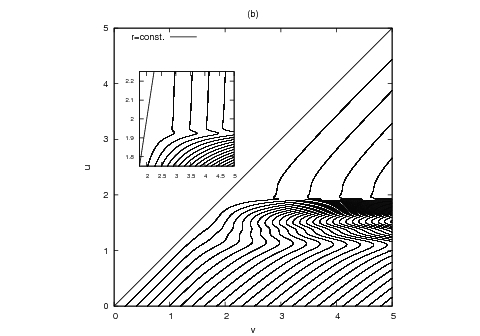}
\caption{Lines of constant $r$ in $(vu)$-plane for: (a) a slightly curved and (b) an almost critical non-singular 
spacetime left after the evolution of the discussed fields. 
The values of free parameters and coupling constants are the same as in Fig.\ref{fig6}.}
\label{fig10}
\end{figure}
\begin{figure}[p]
\includegraphics[scale=1.25]{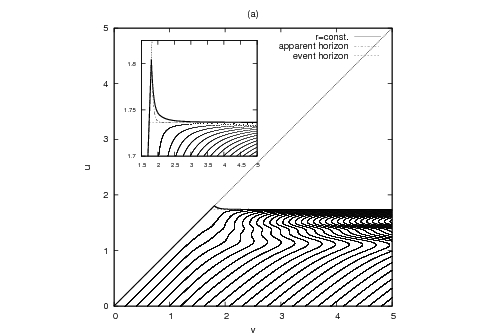}
\includegraphics[scale=1.25]{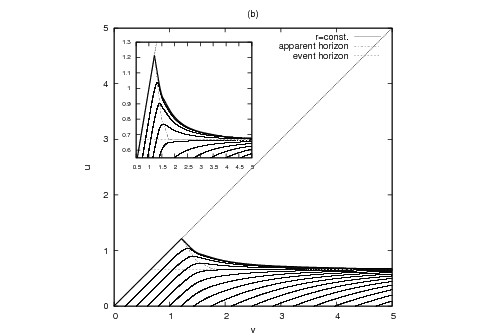}
\caption{Lines of constant $r$ in $(vu)$-plane for spacetime containing: 
(a) a small and (b) a large black hole emerged from the considered gravitational 
collapse. Formation of an apparent and an event horizon and the position of singularity are shown. 
The values of the free parameters and the coupling constants are the same as in Fig.\ref{fig7}.}
\label{fig11}
\end{figure}
\begin{figure}[p]
\includegraphics[scale=1.25]{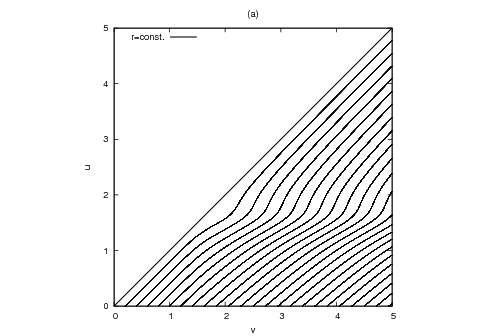}
\includegraphics[scale=1.25]{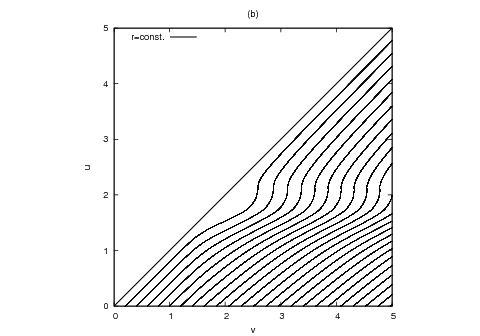}
\caption{Lines of constant $r$ in $(vu)$-plane for: (a) a slightly curved and (b) an almost critical 
non-singular spacetime left after the evolution of the fields in question. 
The values of the free parameters and coupling constants are the same as in Fig.\ref{fig8}.}
\label{fig12}
\end{figure}
\begin{figure}[p]
\includegraphics[scale=1.25]{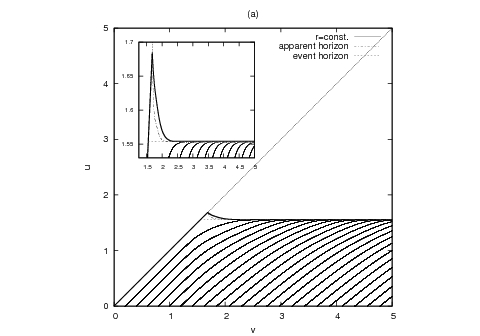}
\includegraphics[scale=1.25]{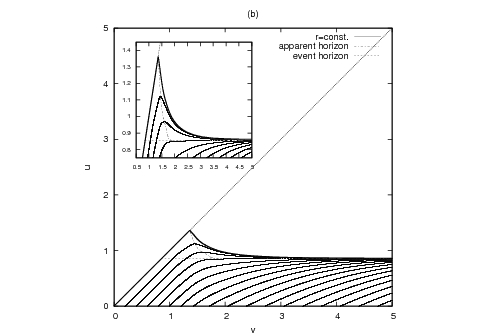}
\caption{Lines of constant $r$ in $(vu)$-plane for spacetime containing: (a) a small and (b) a large black 
hole emerged from the gravitational collapse under consideration. Formation of an apparent and an event horizons
as well as the position of singularity are shown. The values of the free parameters and the coupling constants 
are the same as in Fig.\ref{fig9}.}
\label{fig13}
\end{figure}

In order to better understand the causal structure of the dynamical spacetimes in question 
we shall proceed to perform the Penrose diagrams (the dependence $(u, v)$ along $r = const.$). In Figs.\ref{fig10} and \ref{fig12}
we present the results for the slightly
curved and almost critical non-singular spacetimes for $\alpha = 1$ or $\alpha = -1$ and $e = 0.5$. The other
parameters are as in Fig.\ref{fig6} and \ref{fig8}, respectively. For both of them the outermost contour line corresponding to
$r = 0$ is a non-singular straight line. On the other hand, in Figs.\ref{fig11} and \ref{fig13} we presented lines of
constant $r$ in the $(vu)$-plane for small and large black hole emerging from the gravitational collapse.
Fig.\ref{fig11} was performed for $\alpha = 1$, while in Fig.\ref{fig13} we put the coupling constant $\alpha = - 1$. 
In both cases one has $e = 0.5$.
The outermost thick line is equivalent to $r = 0$. Contrary to the previous cases, in the 
spacetime of dynamical formation of black hole one has a straight line ($u = v$) in the left section.
This behaviour corresponds to the non-singular axis. On the other hand, the right part corresponds to the central
singularity at $r = 0$. Because of the fact that $r_{,v} < 0$ along the latter section, we have to do
with the spacelike singularity. Two types of horizons are present in the singular spacetimes. The apparent horizon
is represented by the contour $r_{,v} = 0$, while the event horizon is provided by the line of constant $u$ and
characterized by $r_{,v} = 0$ when $v \rightarrow \infty$. The dynamical character of the emerging
spacetimes is reflected in relative positions of the aforementioned horizons. They do not coincide  
in the early stages of the evolution and in the end, they  approach the same value of $u = const.$ as $v$ tends to
infinity.

\par
\begin{figure}[p]
\includegraphics[scale=0.7]{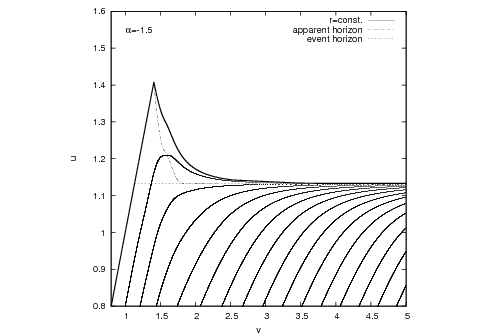}\includegraphics[scale=0.7]{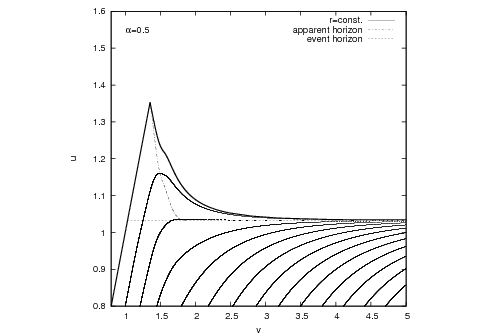}
\includegraphics[scale=0.7]{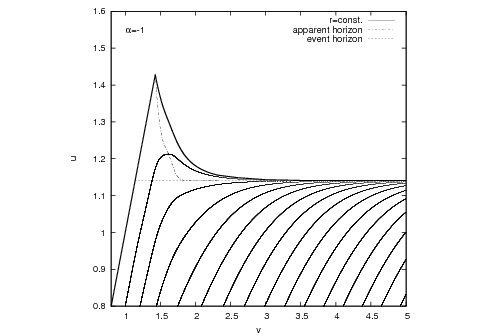}\includegraphics[scale=0.7]{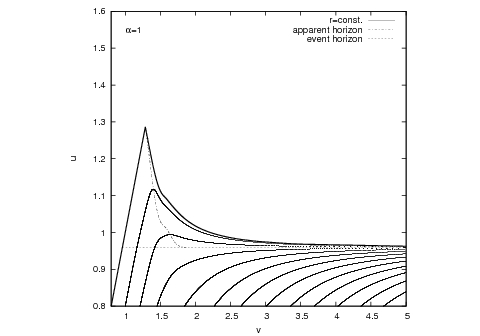}
\includegraphics[scale=0.7]{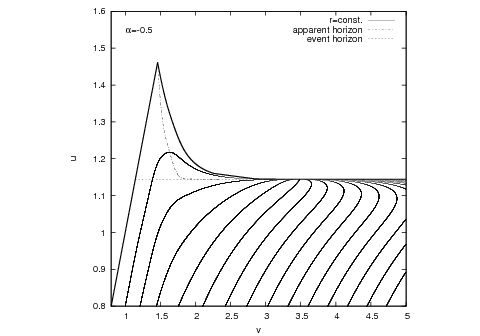}\includegraphics[scale=0.7]{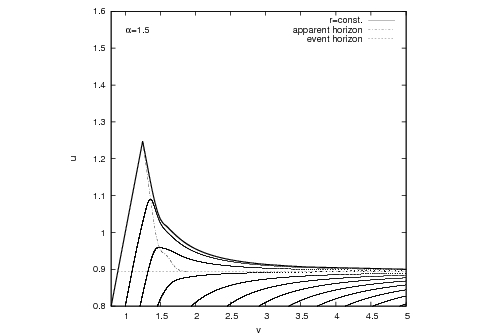}
\caption{Lines of constant $r$ in $(vu)$-plane for various values of coupling constant $\alpha$.}
\label{fig14}
\end{figure}

The next object of our interest was an influence of dilatonic coupling constant on the 
evolution described by the considered equations of motion.
The set of Penrose diagrams (representing lines $r=const.$ in $(vu)$-plane) for different values 
of $\alpha$ is shown in Fig.\ref{fig14}. The values of dilatonic coupling constants 
were arbitrarily chosen to be $\pm 0.5$, $\pm 1$ and $\pm 1.5$. The value of the free family parameter 
was taken as $\tpe=0.075$. We conclude that the value of $\alpha$ exerts no qualitative 
influence on the structure of spacetime. The most striking effect 
is connected with the moment of the horizon's formation. In general, 
for bigger absolute values of $\alpha$ the horizon forms at earlier advanced times. Although 
the effect is far more noticeable for positive values of dilatonic coupling constant, 
it is also present for $\alpha$ not exceeding zero.

\begin{figure}[p]
\includegraphics[scale=1.25]{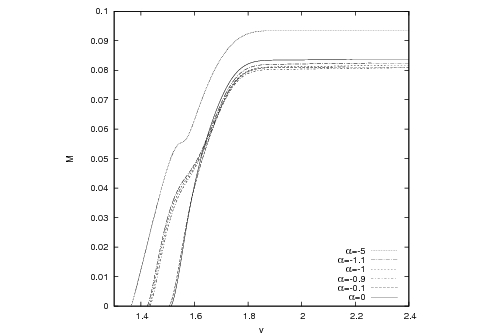}
\caption{Mass of a black hole as a function of $v$-coordinate along the apparent horizon for different negative values of coupling constant $\alpha$.}
\label{fig15}
\end{figure}

\begin{figure}[p]
\includegraphics[scale=1.25]{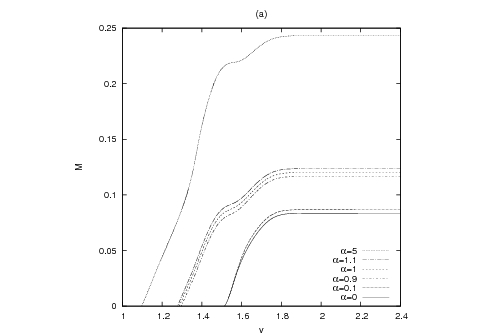}
\includegraphics[scale=1.25]{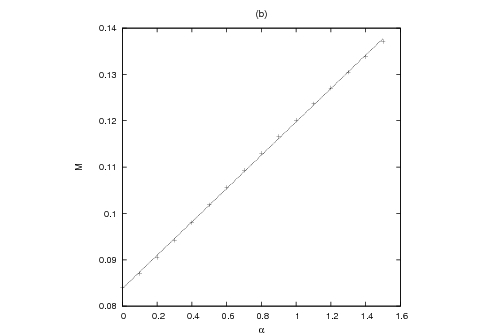}
\caption{The influence of positive values of coupling constant $\alpha$ on 
mass of emerging black hole: (a) mass as a function of $v$-coordinate 
along the apparent horizon and (b) an asymptotic value of black hole mass 
(for $v=3$) as a function of coupling constant $\alpha$.}
\label{fig16}
\end{figure}

We also examined mass of a black hole emerging from the gravitational collapse in question 
as a function of $v$-coordinate along the apparent horizon for different values of coupling constant $\alpha$. 
The relations are depicted in Figs.\ref{fig15} and \ref{fig16}. Mass of a black hole, 
denoted by $M$, is Hawking mass (\ref{mhaw}) calculated along apparent horizon. It turned out that for 
large values of retarded time black hole mass tends to a constant value. For positive values of $\alpha$ 
we observe that the bigger
coupling constant is the bigger Hawking mass one gets. Moreover,
the dependence of $M$ on $\alpha$-coupling constant is linear.
In the case of $\alpha$ less than zero,
its influence on the asymptotic value of Hawking mass is not so straightforward.

\par
Now we proceed to study
different features of the collapsing spacetimes as functions of null coordinates.
Namely,
we shall concentrate on Hawking mass (\ref{mhaw}), Ricci scalar (\ref{eqn:Ricci-scalar-u-v-1}), metric coefficient $g_{uv}$ and 
$v$-derivative of $r$. In 
our considerations we put $\tpe=0.1$, $e=0.5$
and considered two distinct types of evolutions for dilatonic coupling constants equal to $\alpha = \pm 1$.
We took into account foliations of spacetime in both $v$- and $u$-directions and
studied ingoing null rays terminating at a non-singular part of $r=0$, outside event horizon 
and within it, as well as at singular $r=0$. We paid attention to outgoing null rays outside, 
inside and exactly along the event horizon of the emerging black hole. 
For both values of $\alpha$ we
examine ingoing null ray $v=0.75$ lying entirely outside the event horizon, hence terminating at the non-singular part of $r=0$. 
Next, we consider a null ray situated along $v=1$ which crosses the event horizon, 
but also ends at regular $r=0$. The other ingoing null rays 
are $v=1.25$ for $\alpha=-1$ and $v=1.2$ for $\alpha=1$. They have the same characteristic as 
the previous one, but they are situated much closer to the cusp on Penrose diagram of the spacetime. 
We also consider two null rays terminating at central singularity which lie along $v=1.75$ and $v=2$.
As far as the outgoing null rays are concerned, one
examined the outgoing null ray $u=0.5$ situated outside event horizon for $\alpha=-1$ and $\alpha=1$. 
The event horizon is situated along $u=0.8557$ for $\alpha=-1$ and along $u=0.6663$ for $\alpha=1$, 
so these null rays were the next objects of our studies. We also paid attention to $u=const.$ lying beyond the event horizon. 
They are situated along $u=0.86$, $u=0.95$ for $\alpha=-1$ and along $u=0.67$, $u=0.75$ for $\alpha=1$.
\par
\begin{figure}[p]
\includegraphics[scale=1.25]{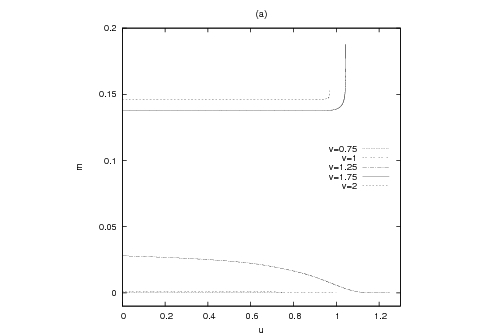}
\includegraphics[scale=1.25]{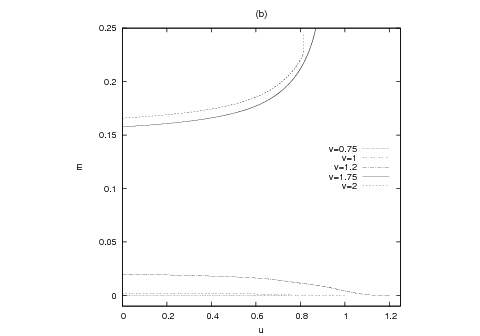}
\caption{Hawking mass as a function of $u$-coordinate along ingoing null rays for: (a) $\alpha=-1$ and (b) $\alpha=1$.}
\label{fig17}
\end{figure}

Hawking mass as a function of $u$ along null rays in question 
for $\alpha=-1$ and $\alpha=1$ is shown in Fig.\ref{fig17}. 
For both values of $\alpha$, in the case of ingoing null rays terminating at non-singular part of $r=0$,
 Hawking mass tends to zero as the retarded time increases. 
Its behaviour along $v=const.$ is evidently different for ingoing null rays hitting the central singularity. 
For $\alpha=-1$ Hawking mass is constant until it reaches the vicinity of $r=0$, 
where it grows rapidly. The situation is
quite different for the coupling constant
$\alpha=1$. In this case Hawking mass grows continuously along $v=const.$ 
and its increase is not so sudden close to $r=0$.
\par
\begin{figure}[p]
\includegraphics[scale=1.25]{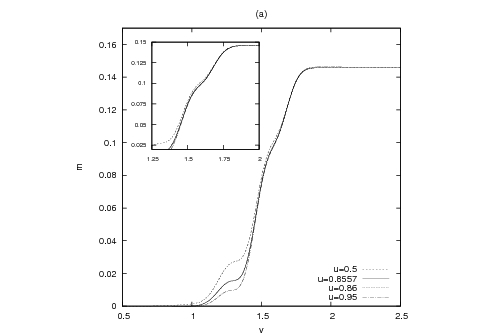}
\includegraphics[scale=1.25]{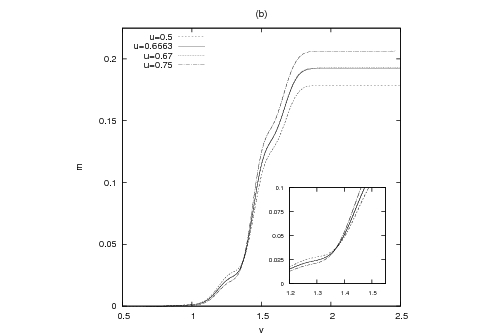}
\caption{Hawking mass as a function of $v$-coordinate along outgoing null rays for: (a) $\alpha=-1$ and (b) $\alpha=1$.}
\label{fig18}
\end{figure}

Hawking mass as a function of $v$ along 
the aforementioned null rays for $\alpha=-1$ and $\alpha=1$ is shown in Fig.\ref{fig18}. 
It grows during the dynamic evolution and reaches a constant value 
when the spacetime becomes static, which is indicated by the coalescence of apparent and event horizons. 
In the early stages of the collapse, for both $\alpha$, 
the values of Hawking mass are bigger for small retarded times. 
For $\alpha=-1$ at some $v$-interval this relationship disappears and Hawking mass randomly reaches actually 
the same constant value for all outgoing null rays. 
The interval is shown in the magnified area in Fig.\ref{fig18}a. 
On the contrary, for $\alpha=1$ there is one point, 
at which lines indicating Hawking mass as a function of $v$ cross. The relationship 
mentioned above changes its character and values of Hawking mass are now bigger for 
large retarded times. The point and its vicinity are depicted in the magnified area in Fig.\ref{fig18}b. 
Moreover, for $\alpha=1$, the constant values of Hawking mass for large advanced times 
are different for each $u=const.$ 
\par
\begin{figure}[p]
\includegraphics[scale=1.25]{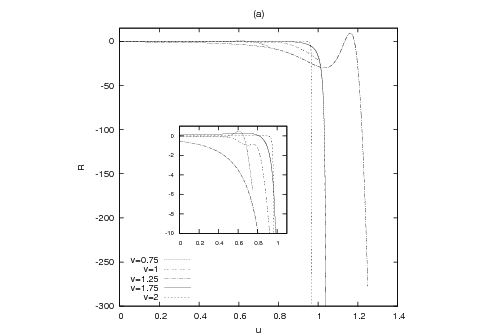}
\includegraphics[scale=1.25]{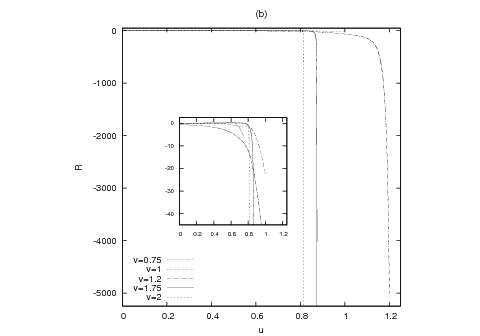}
\caption{Ricci scalar as a function of $u$-coordinate along ingoing null rays for: (a) $\alpha=-1$ and (b) $\alpha=1$.}
\label{fig19}
\end{figure}
\begin{figure}[p]
\includegraphics[scale=1.25]{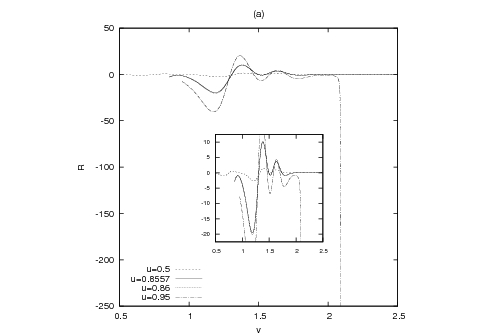}
\includegraphics[scale=1.25]{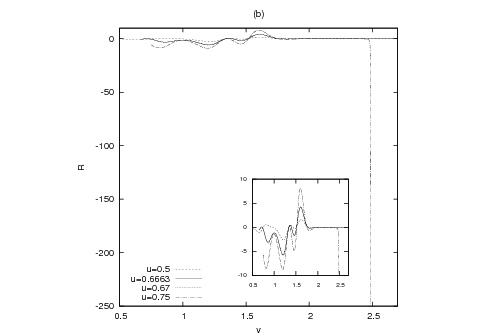}
\caption{Ricci scalar as a function of $v$-coordinate along outgoing null rays for: (a) $\alpha=-1$ and (b) $\alpha=1$.}
\label{fig20}
\end{figure}

Ricci scalar as a function of $u$ and $v$ along the considered null rays is shown in Figs.\ref{fig19} and \ref{fig20}. 
In flat regions of the spacetime its values are practically equal to zero. 
For $\alpha=-1$ and $\alpha=1$ Ricci scalar is finite along ingoing null rays 
terminating at the non-singular $r=0$ and 
along outgoing null rays escaping to infinity. 
It is divergent along null rays ending at the singular part of $r=0$.
\par
\begin{figure}[p]
\includegraphics[scale=1.25]{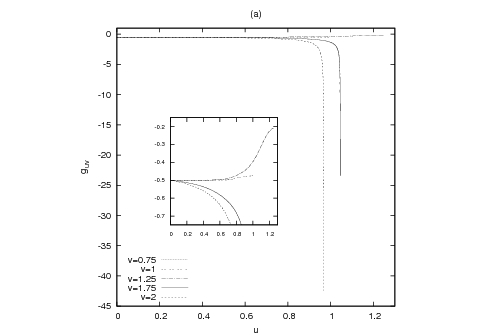}
\includegraphics[scale=1.25]{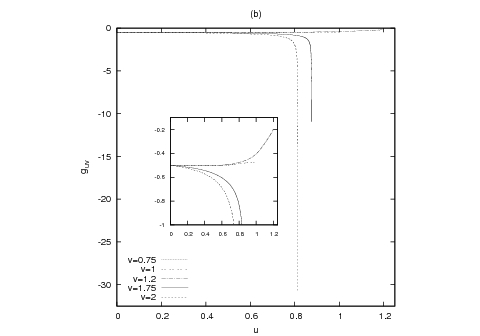}
\caption{Metric coefficient $g_{uv}$ as a function of $u$-coordinate along ingoing null rays for: (a) $\alpha=-1$ and (b) $\alpha=1$.}
\label{fig21}
\end{figure}
\begin{figure}[p]
\includegraphics[scale=1.25]{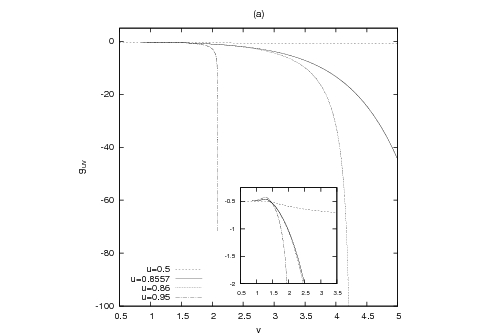}
\includegraphics[scale=1.25]{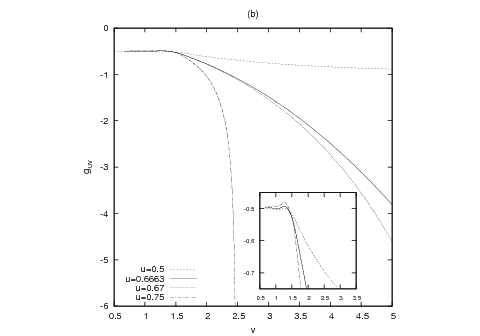}
\caption{Metric coefficient $g_{uv}$ as a function of $v$-coordinate along outgoing null rays for: (a) $\alpha=-1$ and (b) $\alpha=1$.}
\label{fig22}
\end{figure}

On the other hand, the metric coefficient $g_{uv}$ as a function of $u$ and $v$ along 
null rays in question is shown in Figs.\ref{fig21} and \ref{fig22}. 
For both values of $\alpha$ it is constant in flat regions of spacetime. 
In curved areas it is slightly increasing with the retarded time for ingoing null rays terminating at a regular part of $r=0$ 
and considerably decreasing in the other case. $g_{uv}$ has a peak near the value of the advanced time, 
where the apparent horizon appears and the spacetime becomes singular. 
The peaks for both values of $\alpha$ are shown in magnified areas of the respective figures. 
Then, $g_{uv}$ decreases with the advanced time along all outgoing null rays. 
The changes are rather small outside the apparent horizon and become large beyond it.
\par
\begin{figure}[p]
\includegraphics[scale=1.25]{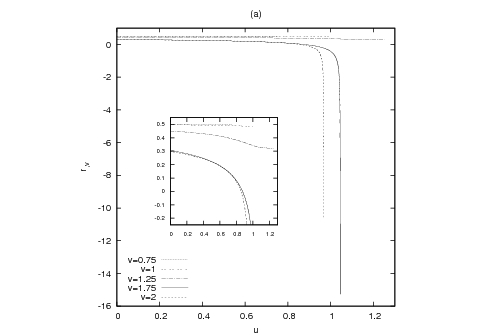}
\includegraphics[scale=1.25]{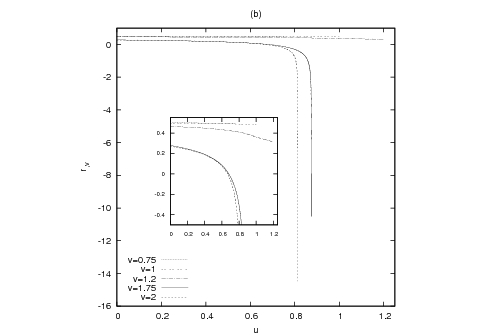}
\caption{Quantity $r_{,v}$ as a function of $u$-coordinate along ingoing null rays for: (a) $\alpha=-1$ and (b) $\alpha=1$.}
\label{fig23}
\end{figure}
\begin{figure}[p]
\includegraphics[scale=1.25]{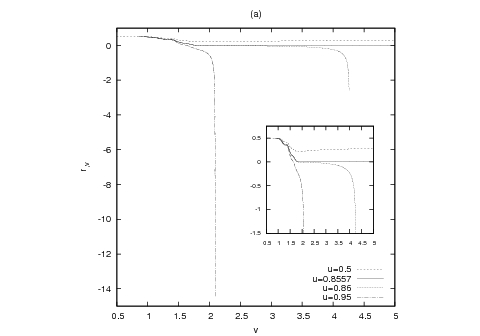}
\includegraphics[scale=1.25]{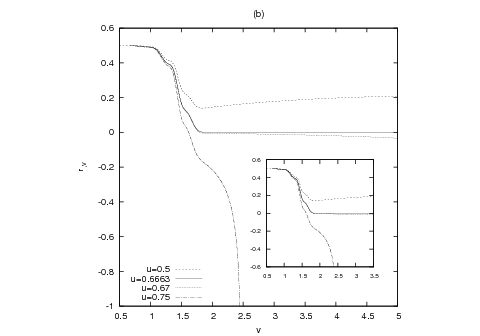}
\caption{Quantity $r_{,v}$ as a function of $v$-coordinate along outgoing null rays for: (a) $\alpha=-1$ and (b) $\alpha=1$.}
\label{fig24}
\end{figure}

The derivative of $r$ with respect to $v$ as a function of $u$ along null rays in question is shown in Fig.\ref{fig23}. 
It is constant or slightly decreasing for ingoing null rays lying outside the apparent horizon 
and displays a strong variability along $v=const.$, which crosses the apparent horizon.
On the other hand, the
derivative of $r$ with respect to $v$ as a function of $v$ along the examined null rays is shown in Fig.\ref{fig24}. 
It is constant in the $v$-interval corresponding to the non-singular part of $r=0$. 
In the region, where the line indicating a central singularity on Penrose diagram is the steepest $r_{,v}$ decreases rapidly, 
but not uniformly - it in turn displays quick and slow changes. 
In the area, where the singular $r=0$ lies almost along a constant $u$ the derivative of $r$ with respect 
to $v$ changes only a little. It increases for outgoing null rays outside the event horizon, is constant along it 
and decreases within it.

\section{Conclusions}
In our paper we considered a dynamical collapse of a charged complex 
scalar field in the low-energy limit of the string theory, the so-called dilaton gravity.
In order to solve the complicated equations of motion one used the double-null coordinates which
enabled us to begin with the regular initial spacetime at null infinity, compute the formation of a black hole           
and broaden our inspection to the singularity formed during the dynamical collapse in question.
We have formulated the problem under consideration as a system of the first order partial differential
equations with the adequate boundary and regularity conditions.
\par
We begin our numerical studies with a subcritical evolution. One observes that there is a flat region 
in $(rv)$-plane corresponding to small values of advanced and retarded coordinates where $r(u, v)$
is proportional to $u$ and $v$-coordinates. Then, the curvature of the spacetime comes into being,                  
destroying this behaviour. The spacetime curved earlier and more significantly for the larger values of
$\tpe$ parameter. Next we proceed to analyze the Kruskal diagrams $(r,v)$ along $u = const.$ for the considered spacetimes.
We found that the outermost null rays escaped to the future null infinity, while the innermost 
originating from the non-singular axis $r = 0$ terminated at the singular part of the hypersurface in question.
We did not observe formation of the inner horizon. The causal structure of the dynamical collapse
was elaborated by means of Penrose diagrams, namely we analyzed $(v,u)$-dependence along $r = const.$
We noticed that the left part of the diagram corresponded to the non-singular axis $r = 0$ and the right section
of it was bounded with the central singularity. We have indicated two types of the horizons, i.e.,
the apparent one represented by the contour $r_{,v} = 0$ and the event horizon forming when $v \rightarrow \infty$.
They did not coincide at the early stages of the dynamical collapse but in the end they approached the same value of 
$u$-coordinate as $v \rightarrow \infty$.
Penrose diagrams for various values of $\alpha$ coupling constant were also found.
\par
On the other hand, we considered Hawking mass as a function of $v$-coordinate along different null rays. It
turned out that it grew during the dynamic collapse and reached the constant value for a static spacetime,
where the coalescence of the apparent and event horizon took place. 
We also analyzed the behaviour of Hawking mass as a function of ingoing null rays terminating at a non-singular part
of $r=0$. It was revealed that it tended to zero with the passage of the retarded time. Hawking mass behaviour,
along $v= const.$ for ingoing null rays hitting the central singularity, was different for the different
values of coupling constant $\alpha$.\\
As far as the Ricci scalar and metric tensor coefficient $g_{uv}$ 
are concerned their behaviour is more or less similar for the positive and negative $\alpha$ coupling constant.
\par
To conclude one remarks that the dynamical collapse of a complex charged scalar field in the realm of dilaton gravity
resembles the collapse leading to the Schwarzschild black hole rather than the collapse of charged field in Einstein-Maxwell theory.
Though, when we check our code and put dilaton field and coupling constant equal to zero we get the behaviour leading
to black hole with an a Cauchy horizon, during dynamical collapse of charged scalar field in dilaton theory
we do not find the inner black hole horizon.

\begin{acknowledgments}
AB was supported by Human Capital Programme of European Social Fund sponsored by European Union.
\end{acknowledgments}






\end{document}